\documentclass[aps,prb,superscriptaddress,papersize=a4paper,twocolumn]{revtex4}
\usepackage{amsmath,amsfonts,amssymb,amsthm}
\usepackage{mathtext,mathtools,mathrsfs, bm, comment}
\usepackage{esint,bm,esvect,dsfont}
\usepackage{graphicx,epsfig,psfrag,xcolor}
\usepackage{array,tabularx,tabulary,booktabs,multirow}
\usepackage{hyperref}
\usepackage{verbatim}

\begin{document}

\title{Anomalous superconductivity and unusual normal state properties of bilayer and twisted graphene (Brief review)}
	
\author{M.~Yu.~Kagan}
\affiliation{National Research University Higher School of Economics, Moscow, 101000 Russia}
\affiliation{Kapitza Institute for Physical Problems, Russian Academy of Sciences, Moscow, 119334 Russia}
\author{M.~ M.~Korovushkin}
\affiliation{Kirensky Institute of Physics, Federal Research Center KSC, Siberian Branch, Russian Academy of Sciences, Krasnoyarsk, 660036 Russia}
\author{V.~A.~Mitskan}
\affiliation{Kirensky Institute of Physics, Federal Research Center KSC, Siberian Branch, Russian Academy of Sciences, Krasnoyarsk, 660036 Russia}
\author{K.~I.~Kugel}
\affiliation{Institute for Theoretical and  Applied Electrodynamics, Russian Academy of Sciences, Moscow, 125412 Russia}
\affiliation{National Research University Higher School of Economics, Moscow 101000, Russia}
 \author{A.~L.~Rakhmanov}
\affiliation{Institute for Theoretical and  Applied Electrodynamics, Russian Academy of Sciences, Moscow, 125412 Russia}
\affiliation{Dukhov Research Institute of Automatics, Moscow, 127055 Russia}
\author{A.~V.~Rozhkov}
\affiliation{Institute for Theoretical and  Applied Electrodynamics, Russian Academy of Sciences, Moscow, 125412 Russia}
\author{A.~O.~Sboychakov}
\affiliation{Institute for Theoretical and  Applied Electrodynamics, Russian Academy of Sciences, Moscow, 125412 Russia}	

\date{\today}

\begin{abstract}
It has been shown that the Kohn--Luttinger superconductivity mechanism interplaying with other types of ordering can be implemented in systems with a hexagonal lattice. A number of unusual properties of such systems in the normal phase have also been considered. Our previous results on Kohn--Luttinger superconductivity with $p$-, $d$-, and $f$-wave pairing in monolayer and AB bilayer graphene, obtained disregarding the effect of substrate potential and impurities, have been presented in the first part. Then, the interplay of the superconducting Kohn--Luttinger state with the spin density wave state in actual AB, AA, and twisted bilayer graphene has been discussed in detail. In the last parts, a number of anomalous properties in the normal phase and the appearance of nematic superconductivity alongside with the spin density wave in the twisted bilayer graphene have been presented.
\end{abstract}

\maketitle

\section{Introduction}
\label{intro}

Since its discovery, graphene is attracting a widespread attention due to its unusual electronic characteristics~
\cite{NovoselovSci2004,KatsnelsonNatPh2006,KotovRMP2012,NetoRMP2009,RozhkovPhRep2016,
SboychakovPRB2015,KaganKugRakhPhRep2021,BookPS}. Experimental studies of bilayer graphene and the discovery of superconductivity in magic angle twisted bilayer graphene in 2018 \cite{CaoNat2018} stimulated an advent of numerous theoretical works on anomalous (non-phonon) mechanisms of superconductivity in graphene and related compounds (see review ~\cite{KaganUFN2015} and references therein), including the Kohn--Luttinger mechanism \cite{KohnLuttPRL1965,FayPRL1968,KaganPZhETF1988,
KaganPZhETF1989,ChubukovJPCM1989,KaganPhLA1991}.

The possibility of implementing anomalous mechanisms of superconductivity in graphene is primarily related to strong electron correlations in this system and is largely due to the existence of Dirac points in the spectrum of graphene. Near these points, the energy spectrum is linear and electrons can be described by the Dirac equation for massless charged quasiparticles. This results in a pronounced electron-polaron effects \cite{KaganKugRakhPhRep2021,BookPS}, which strongly enhances electron correlations. Such fact gives hope for implementing in graphene superconductivity mechanisms based on the electron--electron interaction.

\begin{figure}
    \centering
    \includegraphics[width=0.7\columnwidth]{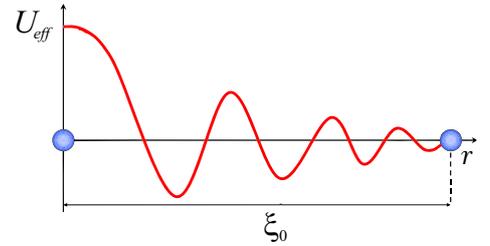}
    \caption{Friedel oscillations in the effective interaction between two particles due to the polarization of
the Fermi background, where $\xi_0$ is the coherence length of a Cooper pair~\cite{KaganUFN2015,BaranovIJMPB1992,KaganBookSpringer}.
\label{fig1}
}
\end{figure}

Note that the Kohn--Luttinger mechanism is the basic one for the non-phonon superconductivity in strongly correlated electron systems with repulsion, and it competes with the Fr\"ohlich plasmon mechanism~\cite{FrohlichJPhC1968} in the systems with low electron density. For the Kohn--Luttinger superconductivity mechanism, the frequency effects associated with the retardation do not play a dominant role, while the presence of the Kohn anomaly~\cite{KohnPRL1959} (Friedel oscillations~\cite{FriedelNuCim1958}) in the effective interaction between two electrons of the Cooper pair via the polarization of the Fermi background plays a crucial role (see Fig.~\ref{fig1}). In fact, the Kohn--Luttinger mechanism involves the transformation of the bare repulsive interaction between two particles in vacuum in the presence of a Fermi background (a filled Fermi sphere) into an effective attraction in the material within a channel characterized by a nonzero orbital moment of the Cooper pair.

Our previous results on the Kohn--Luttinger mechanism~\cite{BaranovIJMPB1992,BaranovZsPhB1992,BaranovPhC1993,ChubukovPRB1993,KaganUFN1994,
BaranovPZhETF1996,BaranovZhETF1991,BaranovZhETF1992,
KaganPZhETF1995,BaranovPZhETF1993} are presented in the review~\cite{KaganUFN2015} on anomalous superconductivity in electron systems with repulsion. They are dealing primarily with low-density electron systems described by the Hubbard model or by the Fermi gas model with the purely repulsive bare interaction. In the effective interaction of two particles mediated by the polarization of the Fermi background in all channels with nonzero momenta of the Cooper pair, the attraction arises, and the most attractive harmonic corresponds to the triplet $p$-wave pairing~\cite{FayPRL1968,KaganPZhETF1988}. Moreover, the critical temperature corresponding to the $p$-wave pairing can be significantly enhanced in the spin-polarized Fermi gas or in the two-band situation, leading to reasonable values of $T_c = 1-5$ K~\cite{KaganPZhETF1989,KaganPhLA1991, BaranovIJMPB1992,BaranovPhC1993,KaganUFN1994}. In doing so, there appears a possibility to enhance the Kohn anomaly in the three-dimensional case. However, it is more important that in the case when the energy band becomes almost flat, it is possible to make the strong but one-sided (square-root type) Kohn anomaly efficient for the superconductivity problem already in the second order of perturbation theory. Note that, according to the results reported in \cite{KaganUFN2015,BaranovIJMPB1992,KaganBookSpringer}, the Kohn anomaly plays an important but not dominant role in the implementation of the $p$-wave pairing. In a more general case, it is sufficient for the effective (or screened) interaction to increase only in the range from 0 to 2$p_F$. Note also that in repulsive models on a two-dimensional square lattice, such as the Hubbard model with repulsion, the $d_{xy}$-wave pairing naturally arises in a wide electron density range, whereas at high electron densities (closer to half-filling), the $d_{x^2-y^2}$-wave pairing, relevant for cuprate superconductors, takes place~\cite{BaranovIJMPB1992,KaganBookSpringer,BaranovZsPhB1992}.

\section{Superconductivity in monolayer graphene and in idealized bilayer graphene}
\label{idealized BLG}

\begin{figure}
    \centering
    \includegraphics[width=0.8\columnwidth]{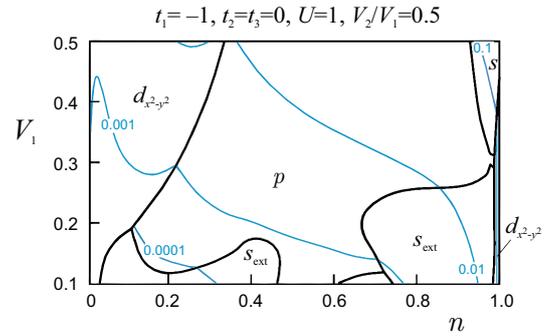}
    \caption{(Color online) $n-V_1$ phase diagram of the Shubin--Vonsovsky model on the square lattice at
$t_2=t_3=0$, $U=|t_1|$ and $V_2/V_1=0.5$. Thin lines correspond to fixed values of the coupling constant  $|\lambda|$~\cite{KaganZhETF2013}.
\label{fig2}
}
\end{figure}

More recent results of our group on the Kohn--Luttinger mechanism on the lattice~\cite{KaganKugRakhPhRep2021,BookPS,KaganUFN2015,KaganZhETF2011,
KaganPZhETF2011,KaganPZhETF2013,KaganZhETF2013,
KaganZhETF2014,KaganSSC2014,KaganZhETF2014a,KaganEPJB2015,
KaganPZhETF2016} in cuprate and iron-based superconductors, as well as in the monolayer and idealized AB bilayer graphene pertain to the period of 2011‒2016 (see also the results reported in ~\cite{NandkishoreNatPh2012,NandkishorePRL2012,NandkishorePRB2012,
NandkishorePRB2014,DongPRB2023}). In these works, phase diagrams of the superconducting state for cuprate and iron-based superconductors were constructed within the framework of the most repulsive basic Shubin--Vonsovsky model~\cite{ShubVonsPrRoySoc1934} (or extended Hubbard model) taking into account the onsite Hubbard repulsion $U$ and the additional Coulomb repulsion between electrons at different sites of the square lattice, as well as the long-range intersite hopping, generalizing our previous results on $p$- and $d$-wave pairing in the Hubbard model~\cite{KaganPZhETF2013,KaganZhETF2013} (see Fig.~~\ref{fig2}).

\begin{figure}
    \centering
    \includegraphics[width=0.8\columnwidth]{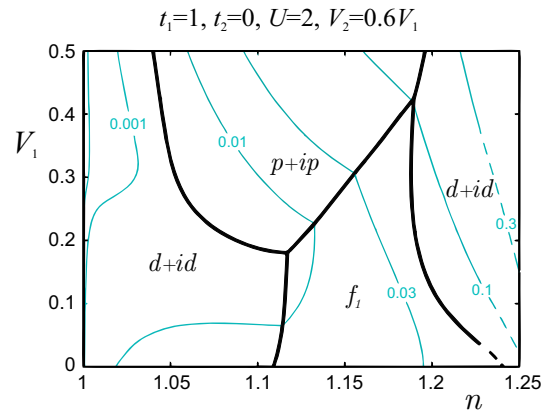}
    \caption{ $n-V_1$ phase diagram for the superconducting state of the graphene monolayer at  $U = 2$ and $V_2 = 0.6V_1$ (all parameter are given in units of $|t|$)~\cite{KaganUFN2015,KaganZhETF2014,KaganSSC2014}.
\label{fig3}
}
\end{figure}

The results of the studies on the Kohn--Luttinger mechanism on a square lattice were also generalized by us to the hexagonal lattice of graphene. In \cite{KaganZhETF2014,KaganSSC2014}, in the framework of the Shubin--Vonsovsky model, we constructed a phase diagram for an idealized graphene monolayer containing, at different electron densities, the regions of triplet pairing with $p_x+ip_y$- and $f$-wave symmetries, as well as small regions of the singlet $d_{x^2-y^2}+id_{xy}$-pairing~\cite{KaganUFN2015,KaganZhETF2014,KaganSSC2014} (see Fig.~\ref{fig3}).

\begin{figure}
    \centering
    \includegraphics[width=0.8\columnwidth]{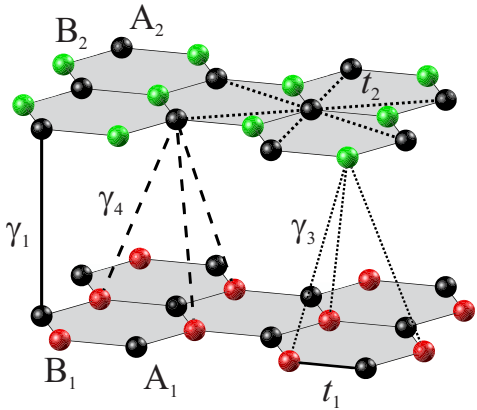}
    \caption{Crystal structure of a AB bilayer graphene, where A and B denote sublattices in the layers. Here,  $t$ and $\gamma$ are the intra- and interlayer hopping integrals, which correspond to the Coulomb interaction energies $V$  and $G$, respectively~\cite{KaganUFN2015,KaganZhETF2014a,KaganEPJB2015}.
\label{fig4}
}
\end{figure}

Then, our studies on the Kohn--Luttinger superconductivity mechanism and its possible enhancement in AB bilayer graphene appeared (see Fig. ~\ref{fig4})~\cite{KaganUFN2015,KaganZhETF2014a,KaganEPJB2015}. We have also demonstrated the enhancement of the critical temperature for $f$-wave and for the chiral $d+id$-wave pairing within the framework of the Shubin--Vonsovsky model, taking into account the long-range hoppings and Coulomb intralayer and interlayer repulsion (alongside with the Hubbard interaction). The calculations were performed using the Born approximation in the second order (in terms of the interaction) of perturbation theory in the idealized AB model of bilayer graphene.

\begin{figure}
    \centering
    \includegraphics[width=0.8\columnwidth]{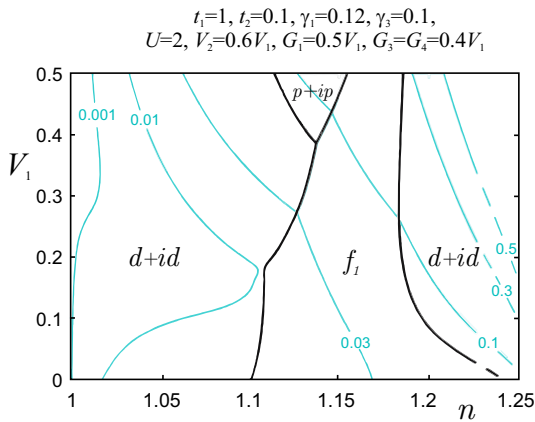}
    \caption{$n-V_1$ phase diagram for the superconducting state of the idealized AB bilayer graphene obtained for the Shubin--Vonsovsky model taking into account long-range intra- and interlayer hoppings and the Coulomb interactions between the nearest and next-nearest neighbor sites~\cite{KaganUFN2015,KaganZhETF2014a,KaganEPJB2015}.
\label{fig5}
}
\end{figure}

The corresponding phase diagram for the superconducting state is shown in Fig.~\ref{fig5}. It exhibits extended regions corresponding to the chiral $d+id$- and $f$-wave pairings (alongside with small regions of the $p+ip$-wave pairing).

Let us emphasize that in the mentioned series of works, we neglected possible interplay of superconductivity with other instabilities in the monolayer and
AB bilayer graphene, and also considered the pure limit in the absence of impurities and defects, to which anisotropic pairing with a nonzero orbital momentum is very sensitive. This implies a natural question concerning the efficiency of the various superconductivity mechanisms in the transition from idealized to actual bilayer graphene.

\section{Competition between superconductivity and spin density wave state in actual AB and AA graphene bilayers, as well as in twisted graphene}
\label{competition SC-SDW}

Note that a significant progress in the study of anomalous superconductivity mechanisms, and, in particular the Kohn‒Luttinger mechanism in various
modifications of actual (not idealized) bilayer graphene, as well as in the currently very popular twisted graphene, is to a certain extent related to the
recent studies of the authors of this review~\cite{RozhkovPhRep2016,SboychakovPRB2015,KaganKugRakhPhRep2021,BookPS,RakhmanovPRL2012,SboychakovPRB2023,RozhkovPRB2023,
SboychakovPRB2024,SboychakovPhE2025}.

The Kohn--Luttinger mechanism of superconductivity in actual graphene competes with other instabilities in the system, such as the formation of spin density waves (SDWs). This competition significantly reduces the superconducting transition temperatures in comparison with the situation in idealized bilayers.

Note also that in bilayer graphene and especially in twisted bilayer graphene, it is necessary to analyze other mechanisms of superconductivity, first of all the Eliashberg mechanism for strong electron--phonon coupling~\cite{BookPS}, as well as more exotic mechanisms such as the BCS--BEC crossover between extended and local Cooper pairs~\cite{CaoNat2018}.

\subsection{Experiment}
\label{exper}

The experimental observation of superconductivity in a graphene-based system was first described in 2018 in \cite{CaoNat2018}, where the parameter range, within which a superconducting state arose in twisted bilayer graphene was delineated. The superconducting twisted bilayer graphene sample was characterized by the twist angle $\theta$, which is close to the ``magic'' one $\theta \approx 1^\circ$, and superconductivity appeared in it only at certain doping levels. The corresponding critical temperature $T_c$ was different for different samples, but never exceeded $T_c \approx 1.7$\,K. After 2018, experimental studies aimed at finding out superconductivity in graphene-based systems have become noticeably more intense. Superconducting states were found in rhombohedral trilayer graphene~\cite{ZhouNat2021}, as well as in trilayer graphene with the moir\'e superstructure~\cite{ParkNat2021,ChenNat2019}. In 2022, the discovery of superconductivity in AB bilayer graphene was reported in~\cite{ZhouSci2022}. The critical temperature in AB graphene (in contrast to preliminary optimistic estimates) in that experiment was as low as 30 mK. Moreover, to stabilize superconductivity, it was necessary not only to dope the sample, but also to apply an electric or magnetic field. The latter fact clearly suggested an unusual order parameter and strong interplay of different instabilities in actual (not idealized, as in \cite{KaganZhETF2014a,KaganEPJB2015}) AB graphene.

\subsection{Theory}

While discussing the Kohn--Luttinger mechanism of superconductivity~\cite{KohnLuttPRL1965} or any other one, in actual graphene-based systems, it is worth to recall that superconductivity in such materials will compete or coexist with an insulating ordered state (e.g., with SDW or its analog). The conditions of this competition/coexistence will necessarily affect the manifestations of the superconducting mechanism, as well as its phenomenological characteristics. To illustrate this idea, we consider the scenario of emergence of superconductivity in AB bilayer graphene discussed in~\cite{SboychakovPRB2023}. As a starting theoretical point, here we justify the argument that undoped AB graphene is a so-called ``layer antiferromagnet''. This is an ordered many-particle state being a specific kind of SDW, in which each layer has a finite spin magnetization, but the magnetizations of the layers are mutually opposite cancelling each other.

The study of both superconductivity and SDW phases is based in \cite{RozhkovPRB2023} on the mean-field approach (in the $T = 0$ limit such an approach does not contradict the Mermin--Wagner--Hohenberg theorem, which is formulated only for $T > 0$). Another important theoretical assumption is the applicability of the random phase approximation (RPA) to determine the screened (effective) Coulomb interaction. This approximation is often criticized as  uncontrolled, but for graphene-based systems, its applicability is usually justified by the fact that the additional degeneracy with respect to the valley quantum number leads to an increase in the factor at the RPA diagrams from two (due to the spin degeneracy) to four (due to spin and valleys).

As the first step, it is necessary to calculate numerically the screened interaction between electrons in bilayer graphene in the RPA. The calculations should be performed both for the intralayer and interlayer interactions. The next step is to choose the order parameter and derive the effective Hamiltonian in the mean-field approximation. Then, the free energy of the system is calculated, minimizing which one can find the value of the order parameter. Within the framework of this formalism, the SDW order parameter in undoped AB graphene was calculated to be $\Delta_{\rm SDW} \approx 4.9$ meV, which agrees quite well with the experimental data on the transport gap~\cite{VeliguraPRB2012}.

This estimate makes it clear that without the ``outside assistance'', superconductivity cannot successfully compete with such a state. Indeed, the characteristic SDW energy scale far exceeds 30 mK, which corresponds to superconductivity. In addition, SDW in AB graphene is an insulating phase, which implies the complete absence of one-electron states at the Fermi level. This excludes the possibility of the two-phase coexistence.

\begin{figure}
    \centering
    \includegraphics[width=0.8\columnwidth]{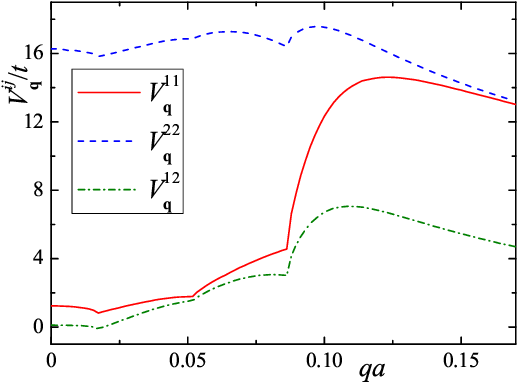}
    \caption{Screened interaction in AB bilayer  graphene $V^{ij}_{\bf q}$, where $i, j = 1, 2$ are layer numbers and ${\bf q}$ is the momentum transfer versus the dimensionless parameter $qa$, where $a$ is the lattice parameter and $q = |{\bf q}|$. The three kinks in the plots are Kohn anomalies (three anomalies instead of the usually observed one are due to the complicated structure of the Fermi surface in the doped system). The interactions $V^{ij}_{\bf q}$ are strongly suppressed at low $q$ values because of the pronounced screening. Superconductivity is due to the (nonmonotonic) increase in the interaction at low momentum transfers.
\label{fig6}
}
\end{figure}

To justify the SDW suppression and the subsequent stabilization of superconductivity, this situation was studied in \cite{SboychakovPRB2023}for the case of electric field applied to the doped AB graphene and directed along the normal to the sample plane. Such field leads to the destruction of the SDW and gives rise to the formation of Fermi surface near the van Hove anomaly. The existence of the doping-induced Fermi surface drastically changes the screening of the Coulomb potential. The RPA calculation demonstrates that the effective interaction is strongly suppressed at low momentum transfers, and the Kohn anomalies become visible \cite{KohnPRL1959}. The plots illustrating the interaction as a function of momentum transfer are shown in Fig.~\ref{fig6}.

In such a situation, is it possible to stabilize superconductivity without phonons, solely due to the Coulomb interaction? Calculations in the mean-field
approximation for the triplet order parameter demonstrate that it is indeed the case. The estimate of the critical temperature $T_c \approx 23$ mK given in \cite{SboychakovPRB2023} agrees well with the available experimental data~\cite{ZhouSci2022}. It is also worth mentioning that the triplet structure of the order parameter obtained using this approach suggests a pronounced stability of superconductivity with respect to the Zeeman field (the magnetic field parallel to the sample plane). The existing experimental data~\cite{ZhouSci2022}, indirectly confirm the fact of such stability.

The authors of \cite{RozhkovPRB2023} pointed out the relation of the calculation technique proposed by them to determine superconducting properties to the Kohn--Luttinger mechanism~\cite{KohnLuttPRL1965}. In particular, they emphasized that, although perturbation theory similar to the classical Kohn--Luttinger formulation do not appear in \cite{SboychakovPRB2024}, the role of the polarization operator, which arises naturally in the RPA, is fundamental for the stabilization of the superconducting order parameter in actual AB bilayer graphene. Just involvement of the polarization operator leads to the suppression of the interaction at low momentum transfers, see Fig.~\ref{fig6}. It directly follows from the calculations that this clearly pronounced nonmonotonic form of the momentum dependence of the interaction underlies the existence of the triplet superconducting order parameter.

Such a program for the analysis of superconductivity was applied in \cite{SboychakovPhE2025} to AA bilayer graphene (see Fig.~\ref{fig7}). It turned out that the  superconductivity mechanism based solely on the Coulomb interaction does not provide any appreciable critical temperature. This is due to the low density of states at the Fermi surface of the system under study. Comparing, for example, AA graphene with AB graphene, discussed above, one
can find that for bilayer graphene with the AB stacking, the density of states grows under effect of doping and electric field tuning, which provide appearing the van Hove singularity at the Fermi level. In the model of AA graphene studied in \cite{ZhouNat2021}, no such possibility allowing one to control the density of states was found.

Concluding the discussion of AA graphene, we would like to remind that this type of bilayer graphene is probably the least studied~\cite{RozhkovPhRep2016}. Despite the availability of samples~\cite{BorysiukJAP2011,LeeJChP2008,LiuPRL2009,RoySurSci1998,GrubisicFrNano2024} and theoretical studies~\cite{RakhmanovPRL2012,RozhkovPRB2023,BreyPRB2013,FarkadMaTodCom2022}, the popularity of AB graphene remains a pipe dream for its AA counterpart.

\section{Unusual characteristics of the normal state of twisted bilayer graphene}
\label{normal TBLG}

\begin{figure}
    \centering
    \includegraphics[width=0.7\columnwidth]{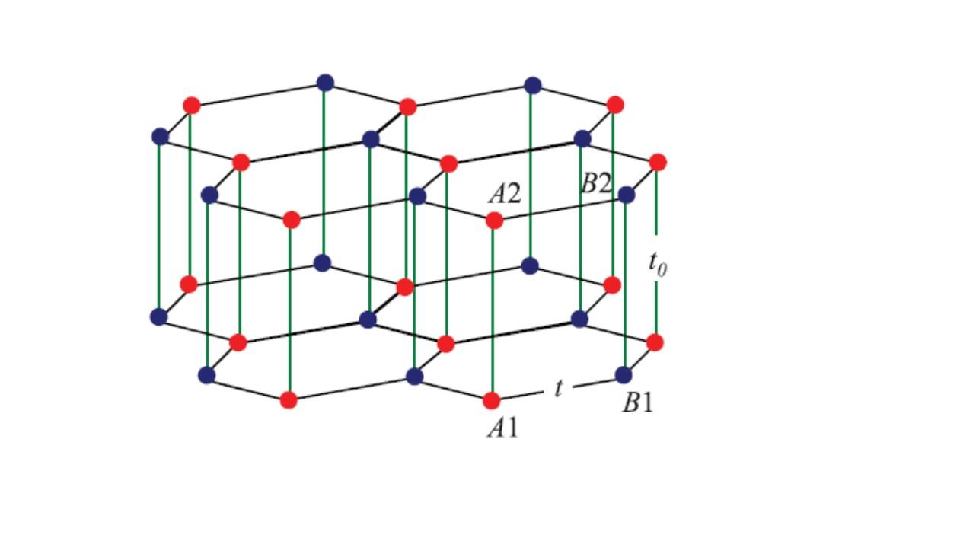}
    \caption{Crystal structure of AA bilayer graphene including (red circles) A and (blue circles) B sublattices. In this crystal structure, the unit cell includes the A1 and B1 atoms in the bottom (1) layer and the A2 and B2 atoms in the top (2) layer.
\label{fig7}
}
\end{figure}

\begin{figure}
    \centering
    \includegraphics[width=0.7\columnwidth]{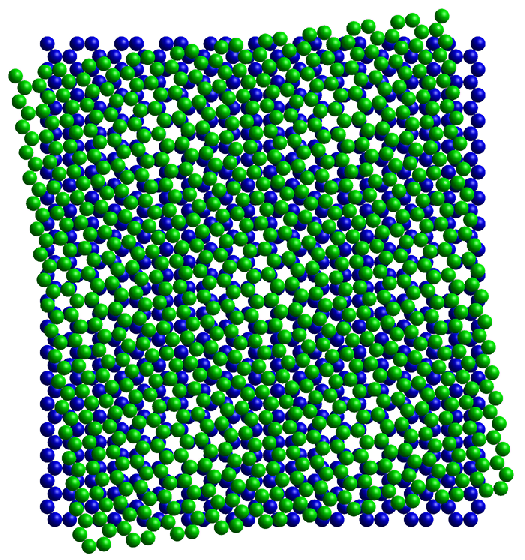}
    \caption{Crystal structure of twisted bilayer graphene at the rotation of one layer by an angle of  $\theta=8^\circ$.
\label{fig8}
}
\end{figure}

In addition to two simple stacking configurations of the bilayer graphene, called AA (see Fig.~\ref{fig7}) and AB (see Fig.~\ref{fig4}), there exists another stacking (or more precisely, a whole set of such configurations), referred to as twisted bilayer graphene (see Fig.~\ref{fig8}). Such material
currently attracts a widespread attention of both theorists and experimentalists.

Note that it is sufficient to rotate the graphene layers relative to each other at a small angle, and we immediately obtain a modulated structure, reminiscent
of moir\'e pattern on some silk fabrics. At small angles of rotation, the unit cell of such moir\'e superlattice can be very large (up to several thousand atoms).

The moir\'e pattern can be visualized by scanning tunneling microscopy (STM)~\cite{LiNatPh2010,YanPRL2012}. By measuring the moir\'e period $L$, one can determine the twist angle $\theta$ according to the formula
\begin{equation}
L=\frac{a}{2\sin(\theta/2)}\,,
\end{equation}
where $a$ is the length of the basis vector of the graphene crystal lattice. The moir\'e pattern appears for an arbitrary rotation, but theorists are mainly dealing with the so-called commensurate angles, for which the system has a periodic superstructure. It can also be shown that in the case of a sufficiently small twist angle, the supercell of twisted graphene can be treated as that consisting of domains with nearly AA, AB, and BA stacking patterns.

The low-energy electronic characteristics of twisted bilayer graphene substantially depend on the twist angle. If this angle is not too small, the low-energy
electron spectrum consists of two doubly degenerate Dirac cones located in two non-equivalent Dirac points of the Brillouin zone of the superlattice.
These Dirac cones intersect at lower and higher energies, forming low-energy van Hove singularities. The Fermi velocity at these Dirac cones is lower than that for graphene monolayer. It decreases with the twist angle. This decrease in Fermi velocity can be significant, but the sample remains to be semimetallic. However, at the so-called first magic angle, the system acquires a Fermi surface even at the charge neutrality point.

\begin{figure}
    \centering
    \includegraphics[width=0.8\columnwidth]{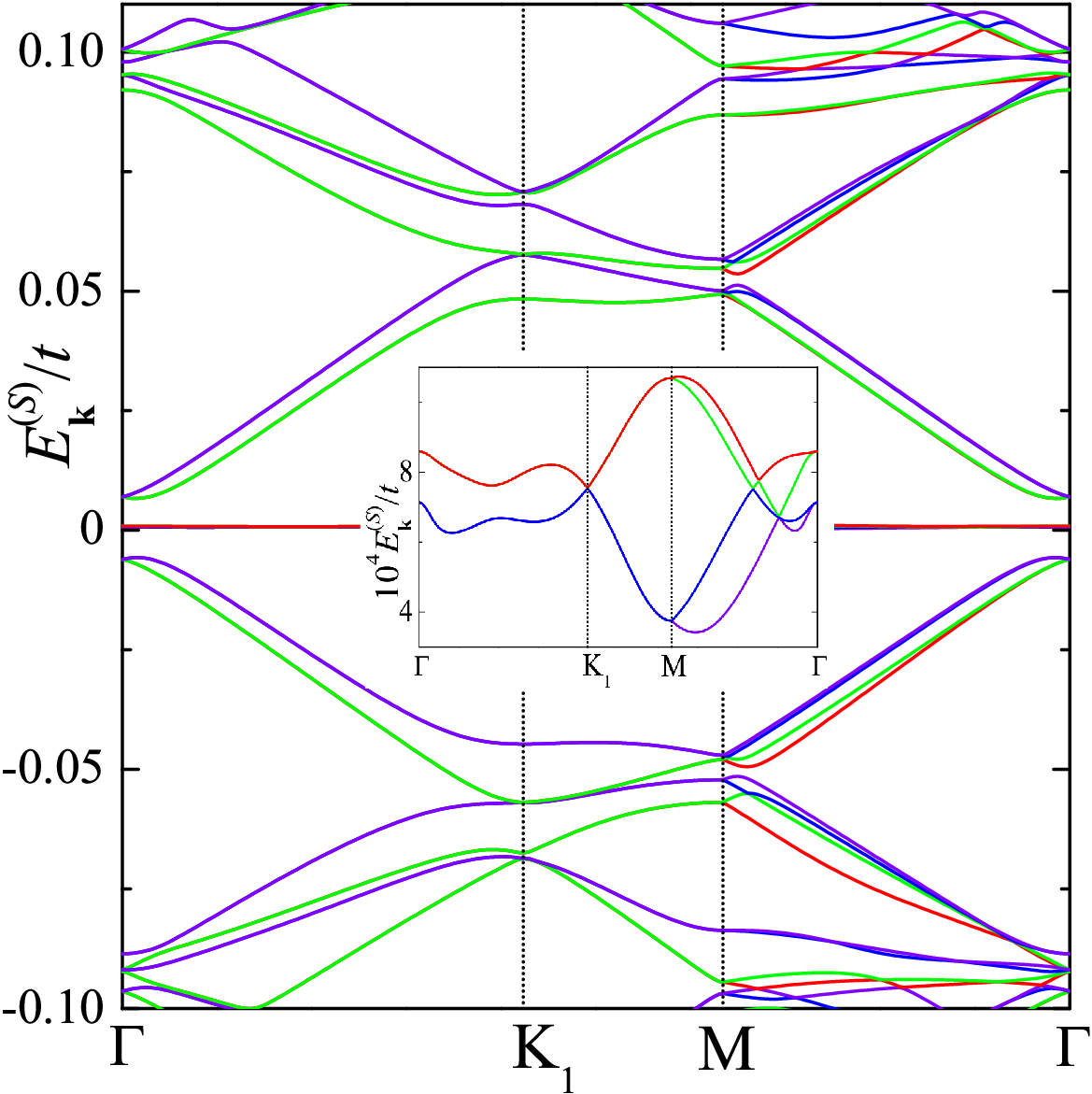}
    \caption{Energy spectrum of twisted bilayer graphene with the superstructure corresponding to the magic angle. Four flat bands are observed near zero energy, separated by band gaps from lower and higher dispersive energy bands with a pronounced dispersion. The inset illustrates the dispersion
curves of the flat bands in the enlarged scale~\cite{KaganKugRakhPhRep2021,BookPS}.
\label{fig9}
}
\end{figure}

The spectrum at the magic angle consists of four almost flat bands (see Fig.~\ref{fig9}), separated from the lower and higher energy bands (with a pronounced dispersion) by band gaps. The width of the flat bands has a minimum at the magic angle. Theoretical predictions and experimental observations show that the first magic angle is as small as $\theta_c \approx 1.08^{\circ}$~\cite{KaganKugRakhPhRep2021,BookPS,CaoNat2018}.

A significant advance in the study of the twisted bilayer graphene was achieved in \cite{CaoNat2018,LiuPRL2018}. The authors of these works managed to develop a technology for manufacturing devices, in which the angle of rotation of the layers can be controlled with an accuracy of tenths of a degree. In these devices, bilayer graphene was placed within a structure that allowed to control the density of charge carriers by applying a gate voltage.

\begin{figure}
    \centering
    \includegraphics[width=0.8\columnwidth]{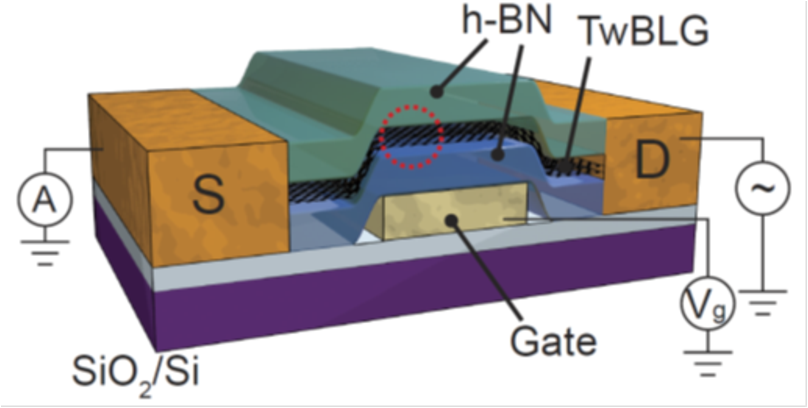}
    \caption{Design of twisted graphene devices used to measure the conductivity as a function of doping~\cite{BookPS,CaoNat2018,LiuPRL2018}.
\label{fig10}
}
\end{figure}

The device under discussion is schematically shown in Fig.~\ref{fig10}. A sample of twisted bilayer graphene with the fixed twist angle is placed between two boron nitride substrates. Two electrodes, source and drain, are located at opposite sides of the sample. A third, gate, electrode is located underneath the sample. Applying a gate voltage provides injection of additional charge carriers, electrons or holes, into the bilayer. This device allows measuring the conductivity of the bilayer depending on the doping level $n$~\cite{CaoNat2018}.

The researchers were interested in the magic twist angles predicted previously by theorists. At the magic angle, the kinetic energy of electrons near the Dirac points is comparable to the energy of interlayer hybridization. The authors measured the conductivity $\sigma(n)$ of several twisted bilayer graphene samples with slightly different twist angles close to the magic one, $\theta_c \approx 1.08^{\circ}$.

In the analysis of experimental data, the researchers were mainly focused on the effects related to the band width. In the case of flat bands, the interaction
becomes very important. Indeed, if the band is quite narrow, then even with a relatively weak Coulomb repulsion of electrons at one part of the lattice, we are dealing with the limit of strong electron correlations. Then, even if the conduction band is half-filled (one electron per unit cell), there appears a gap in the quasiparticle spectrum, since the presence of two electrons within one unit cell is unfavorable owing to their strong Coulomb repulsion. Such a gapped state is a Mott insulator, and it has indeed been observed experimentally at angles close to the magic one. When the effective number of charge carriers is changed by the gate voltage, the gap in the spectrum appears not only for fully occupied minibands (arising due to the moir\'e superlattice), but also for half-filled bands, which is a clear manifestation of the Mott insulator state.

In the experiments, the conductivity plots exhibited minima at band fillings $n = 0, \pm 2, \pm 4$ (some samples also exhibited minima at $n = \pm 3$). The conductivity minima at half-filling, $n = \pm 2$ (i.e., for two additional electrons per supercell) cannot be described within the framework of single-particle theory, and many-particle effects must be taken into account. In addition to the insulating states, the researchers also observed two superconducting ``domes'' slightly below and above the half-filling, $n = -2$.

Note that the interactions break the symmetry of the one-particle Hamiltonian, giving rise to some order parameter in the system. The nature of many-particle insulating states in the magic angle twisted bilayer graphene has been analyzed in many theoretical works. Different types of SDW~\cite{LiuPRL2018,HuangSciBul2019,SboychakovPRB2019,SboychakovPRB2020,SboychakovPZhETF2022,SeoPRL2019}, ferromagnetic state~\cite{CeaPRB2020}, and other states with broken symmetry~ \cite{WuPRL2018,StroppaPRB2021} were proposed as the ground state of the system. It is worth noting that such an abundance of competing insulating phases is not a unique feature of just the twisted bilayer, but is also discussed for other types of bilayer graphene systems, see, for example, \cite{RozhkovSovrEldyn2024}.

The most curious thing is that graphene layers rotated by a magic angle become superconducting at a deviation from the half-filling to one or another side
The maximum superconducting transition temperature $T_c$ observed in experiments is 1.7 K. It seems to be low, but there are few charge carriers in a miniband. Therefore, the ratio of $T_c$ to the Fermi energy $E_{\rm F}$ is higher than that in cuprate superconductors, implying a relative strength of the superconducting pairing. Thus, two main superconducting domes arise in the phase diagram on either side of the Mott insulator state.

In addition to the superconducting domes located near the half-filling $n = -2$, the experimentalists also observed three other superconducting domes near the band fillings $n = 0$ and $n = \pm 1$. Thus, the phase diagram of magic angle twisted bilayer graphene can be very rich. Moreover, the screening of the electron--electron interaction by introducing a metal layer located near the twisted bilayer graphene provides an opportunity to control and tune the shape of the phase diagram.

The electronic characteristics of twisted bilayer graphene can be controlled not only by doping but also by the applied pressure. The applied pressure can be used to control the interparticle distance and, therefore, can affect the magnitude of interaction and the formation of flat bands. For example, the hydrostatic pressure exceeding 11 GPa induced superconductivity in a twisted bilayer graphene device with a twist angle $1.27^{\circ}$, which is larger than the magic value. At zero pressure, such samples did not undergo the transition to the superconducting state.

In high-quality samples of twisted bilayer graphene at twist angles close to the magic value, other ordered phases were found at different numbers of charge carriers per unit cell of the moir\'e superlattice, including some correlated states between the superconducting phases, three of which were insulating the other three, semimetallic. The conducting states were topologically nontrivial.

There was also evidence that near the three-quarter filling level of the conduction miniband, electron correlations favor the transition of twisted bilayer
graphene to the ferromagnetic state. Near this filling level (at 3/4 of the filled band), the system also demonstrates a clearly pronounced anomalous Hall
effect and some signatures of chiral edge states.

\begin{figure}
    \centering
    \includegraphics[width=0.8\columnwidth]{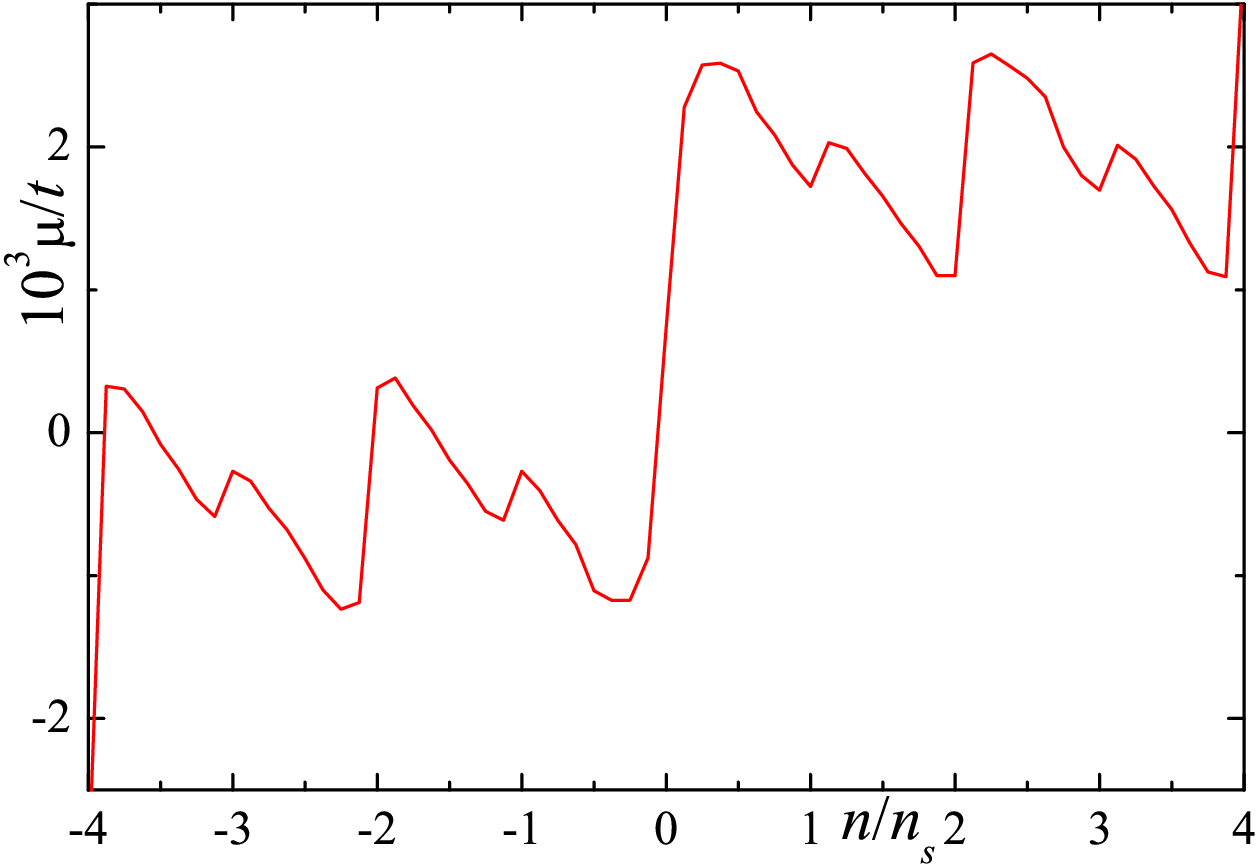}
    \caption{Chemical potential versus doping. The regions with the negative compressibility $\partial\mu(n)/\partial n <  0$ suggest an instability of the homogeneous state with respect to the phase separation~\cite{BookPS,SboychakovPZhETF2020}.
\label{fig11}
}
\end{figure}

The existence of different competing phases usually implies the possibility of phase separation. Indeed, capacitance measurements in samples of twisted
bilayer graphene demonstrate quite nonmonotonic behavior of the electronic compressibility $\partial\mu(n)/\partial n$, where $\mu$ is the chemical potential and $n$ is the number of charge carriers per moir\'e unit cell. The corresponding plots exhibit ranges of negative compressibility (see Fig.~\ref{fig11}), while the  $\mu(n)$ plots include portions with the negative curvature~\cite{BookPS,SboychakovPZhETF2020}. The latter clearly suggests an instability of homogeneous phases and, hence, the electronic phase separation, which is not easy to detect because unit cells in moir\'e superlattices are large.

The theoretical analysis of possible inhomogeneous states in the twisted bilayer graphene near the magic angle was carried out in our work \cite{SboychakovPZhETF2022} (see also \cite{BookPS}). The emerging nonsuperconducting order parameter was assumed to be SDW, and the evolution of such an ordered state with doping was studied. It was shown that in the range of electron densities, where the order parameter is nonzero, the homogeneous state of the system can be unstable with respect to phase separation. It was shown that if the SDW is the ground state of the system, there are four regions of phase separation within the $n < 4$ doping range. Moreover, the phases in the inhomogeneous state are characterized by an even number of electrons per superlattice cell. This allows us to interpret some features in the behavior of conductivity in the doped system. Thus, it becomes possible to explain the fact that the conductivity minima, which could arise at doping levels corresponding to an odd number of electrons per superlattice cell, are absent in some samples under study (the phase separation occurs) and present in other ones (phase separation is suppressed by the long-range Coulomb repulsion).

\section{Nematic superconductivity of twisted bilayer graphene}
\label{nematic SC}

Various superconductivity mechanisms for twisted bilayer graphene (phonons~\cite{WuPRL2018,LianPRL2019}, electron correlations), as well as different symmetries of superconducting order parameters have been discussed in the current literature. In this brief review, we focus on non-phonon mechanisms of superconductivity, such as the Kohn--Luttinger mechanism.

We have already discussed that, if we ignore the interaction and limit ourselves to single-particle calculations~\cite{SboychakovPRB2015}, it turns out that the twisted bilayer graphene exhibits a complicated multicomponent Fermi surface formed by four flat bands appearing at the Fermi level. However, due to the weak dispersion of these bands, this Fermi surface is easily modified due to many-particle effects.

In \cite{SboychakovPRB2024}, the polarization operator and the effective Coulomb interaction were numerically calculated for twisted graphene at different doping levels. This interaction was then used to minimize the many-particle energy in the mean-field approximation. As in the case of the other bilayer graphene species mentioned above, the mean-field calculation demonstrates that the SDW phase is quite stable, as can be expected. Nevertheless, the SDW in twisted graphene is a metallic state rather than an insulating one, since the spin ordering in such system is not able to cover the whole initial Fermi surface by a gap. The survived Fermi surface corresponds to the nematic state, which is confirmed by the experimental data.

Superconductivity can emerge at the residual Fermi surface. The authors of \cite{SboychakovPRB2024} studied the mechanism underlying the formation of this phase based only on the Coulomb interaction. The mean-field calculation demonstrated that the screened Coulomb interaction indeed stabilizes the superconducting order parameter having zeros at the Fermi surface. However, the theoretical estimate for the critical temperature is much lower than the value $T_c \approx 1.7$\, K found in experiments. In \cite{SboychakovPRB2024}, possible causes for this discrepancy are discussed.

Concluding the discussion of \cite{SboychakovPRB2024}], we would like to draw attention to two characteristics of the superconducting phase investigated in it. First, due to the nematicity of the residual Fermi surface, the resulting superconductivity is also nematic. Second, in contrast
to the case of AB graphene, in which the two types of ordering compete with each other, in the twisted graphene, SDW and superconductivity coexist. Such coexistence is a consequence of the inability of  the SDW phase to cover the whole Fermi surface. Thus, we see that superconductivity mechanisms similar to the Kohn--Luttinger scenario can be applied to interpret the experimental situation in bilayer graphene materials of different types.

\section{Conclusions}
\label{concl}

To summarize, the results of our works on the widely discussed issue concerning the implementation of anomalous superconducting pairing in various
modifications of graphene have been reported. The Kohn--Luttinger mechanism of the Cooper instability, which involves the transformation of the initial
repulsive interaction of two particles in vacuum into an effective attraction in the presence of the Fermi background, has been discussed. Our previous results on the possibility of the appearance of the superconducting Kohn--Luttinger phase in idealized monolayer and AB bilayer graphene in the absence of impurities, defects, and van der Waals potential of the substrate, as well as neglecting the interplay with other types of ordering, have been presented. It has been shown that the inclusion of the Kohn--Luttinger renormalizations up to the second order of perturbation theory, as well as the Coulomb interaction of electrons in different carbon atoms of the idealized graphene monolayer, significantly affects the competition between superconducting
phases with $f$-, $d$-, and $p$-wave order parameters. The results obtained for the monolayer have been generalized to the case of an idealized
bilayer graphene having the AB stacking with the interlayer Coulomb repulsion taken into account. It is emphasized that the Kohn--Luttinger mechanism in idealized bilayer graphene systems significantly increases the superconducting transition temperature.

Our recent results on actual AA and AB bilayers, as well as for twisted bilayer graphene, obtained taking into account the interplay between the Kohn--Luttinger superconductivity and the spin density wave, have also been discussed in detail. It has been shown that this interplay sharply reduces the superconducting transition temperatures compared to those in idealized bilayers. In this case, other superconductivity mechanisms, primarily, the electron--phonon Eliashberg mechanism~\cite{BookPS}, can play a decisive role in the development of the Cooper instability. Finally, the appearance of low-temperature nematic superconductivity coexisting with the spin density wave in twisted bilayer graphene near the magic twist angles has been  analyzed.

\section*{Acknowledgments}
We are grateful to V.V. Val’kov and A.V. Chubukov for fruitful discussions.



\end{document}